\documentclass[11pt,a4paper]{article}
\pdfoutput=1
\usepackage{jheppub}
\usepackage{hyperref}

\newcommand\rah{r_{\mt{AH}}}
\newcommand{\be}{\begin{equation}}
\newcommand{\ee}{\end{equation}}
\newcommand{\bea}{\begin{eqnarray}}
\newcommand{\eea}{\end{eqnarray}}

\newcommand{\mt}[1]{\textrm{\tiny #1}}
\def\nc {N_\mt{c}}

\title{Towards Collisions of Inhomogeneous Shockwaves in AdS}
\author[a]{Daniel Fern\'andez}
\affiliation[a]{Crete Center for Theoretical Physics, Department of Physics, University of Crete. PO Box 2208, 71003 Heraklion, Greece}

\date{\today}

\abstract{We perform a numerical simulation of the evolution of inhomogeneities with transverse profile in a collision of gravitational shockwaves in asymptotically anti-de Sitter spacetime. This constitutes a step closer towards an accurate holographic description of the thermalization of a strongly coupled plasma, which can model the dynamics of heavy ion collisions. The results indicate that the considered inhomogeneities typically become hydrodynamical earlier or at the same moment when hydrodynamics applies to the background, even though they decay slowly.
}  
  
\keywords{Gauge-gravity correspondence, Holography and quark-gluon plasmas}

\emailAdd{fernandez@physics.uoc.gr}
 
\begin{document}

\begin{flushright}
CCTP-2014-12 \\
CCQCN-2014-32
\end{flushright}

\maketitle
\setlength{\parskip}{8pt}

\section{Introduction}
\label{intro}

Quantum Chromodynamics (QCD) is a challenging theory in its non-perturbative regime, especially when out-of-equilibrium situations are involved. In particular, the formation of a Quark-Gluon Plasma (QGP) in heavy ion collisions, such as the ones carried out at the RHIC and LHC particle colliders, still lacks a satisfying theoretical description.

The description of the initial stage of this process (right after the collision) would require a full non-perturbative calculation of strongly coupled QCD. Since this is currently beyond our reach, the necessity and importance of developing strong coupling techniques for heavy ion phenomenology has received a boost in recent years. For instance, much work has been done using viscous hydrodynamics \cite{hydro1} or extrapolations from weakly coupled calculations \cite{colorglass}. Although each technique comes with limitations of its own, some successful predictions include those of the shear viscosity to entropy density ratio \cite{starinets} and the elliptic flow coefficient $v_2$ \cite{romatschke1}.

Here we shall take advantage of the convenience of applying the gauge/gravity duality, which allows to access the non-trivial quantum dynamics of this out-of-equilibrium stage by solving classical gravitational dynamics. In this context, our particular holomodel will be constructed for the simplest theory with a gravity dual, ${\cal N}=4$ super Yang-Mills (SYM) theory (in the $\nc, \lambda \rightarrow \infty$ limit) \cite{maldy}. Using this framework, several authors \cite{cy1, yarom, mateos1, mateos2, janikheller, wilkepratt} have previously translated the aforementioned problem into that of a collision of gravitational shock waves in an asymptotically anti-de Sitter (AdS$_5$) spacetime. Such an approach would be equivalent to studying the collision of infinitely extended, homogeneous and planar layers of matter in SYM, which in turn would simulate the highly Lorentz-contracted colliding ions. For an excellent review on this topic, see \cite{wthesis}.

These studies have enlightened ostensibly the longitudinal dynamics involved in heavy ion collisions. However, in the actual experiments, the presence of inhomogeneities and the build-up of momentum in the transverse plane could make transverse dynamics important. In fact, a study including radial flow \cite{wradial} showed that the momentum distribution reaches local equilibrium quickly, after which hydrodynamics applies. Boost-invariance and rotational symmetry were assumed as approximations, in order to keep the numerical calculation effectively 2+1 dimensional\footnote{Indeed, a 3+1 dimensional inclusion of transverse dynamics is non-trivial -- but not impossible, see for instance \cite{3+1} for a remarkable simulation.}. Here, a different formulation will be used to the same effect.

Particularly, as a first approach towards a full calculation including transverse dynamics, we consider the propagation of inhomogeneous perturbations on top of the dynamical gravitational background that encodes the process of thermalization. The dependence on the transverse direction is specified by fixing it to be that of a planar wave. In other words, in each simulation we consider the evolution of a specific Fourier mode.

This problem requires the use of numerical techniques in order to obtain quantitative results. The connection with the dynamics of the plasma is made by extracting the evolution of the stress-energy tensor. Generically, we find that the inhomogeneities decay slowlier than the background thermalizes, which justifies the inclusion of transverse dynamics for a more accurate calculation.
However, the thermalization time, defined according to the applicability of hydrodynamics, is not affected, since the inhomogeneities acquire a hydrodynamic behavior soon enough.

\section{Gravitational description}
\label{grav}
The ansatz for the 5-dimensional spacetime metric is a generalization of that in \cite{cy1}, including additional components to account for the transverse directions $x_1, x_2$ in the form
\bea
\nonumber  & \hspace{-0.4cm} ds^2 = 2dt dr -A\, dt^2  + \Sigma^2 \left[\cosh D\left(e^{B-C}\,dx_1^2+e^{C-2B}\,dy^2\right)+ \right.
\\ & \hspace{-1cm} \left. \sinh D\left(2 e^{B/2} \,dx_1  dy\right)+e^B dx_2^2 \right] + 2dt(F\,dy+G\,dx_1)\, ,
\label{ansatz}
\eea
where $A, B, C, D, \Sigma, F$ and $G$ are functions of the bulk radial coordinate ($r$), time ($t$) and spatial longitudinal ($y$) and transverse ($x_1$, $x_2$) coordinates. The boundary is located at $r=\infty$. Note that the determinant of the spatial part of the metric is a power of a single function, $\Sigma$. This allows to simplify Einstein's equations, by following the characteristic formulation of General Relativity (GR).

For the same reason, we employ generalized ingoing Eddington-Finkelstein coordinates, where paths of varying $r$, with the other coordinates fixed, are infalling radial null geodesics. As a matter of fact, the metric ansatz is invariant under arbitrary reparametrizations of this parameter, $r\to r+\xi(t,y,\vec x)$. This constitutes a gauge freedom that will be fixed by placing the Apparent Horizon (AH) at $r=1$.
The ansatz is complemented with
\be
h(r,t,y,x_1) = h_0(r,t,y)+e^{i k x_1} \delta h(r,t,y)
\label{planewaves}
\ee
where $h$ represents every function in the metric. This is so that the problem simplifies from 3+1 to 2+1 dynamics. The $\delta h$ terms will be treated as perturbations, and Einstein's equations\footnote{with the $AdS$ cosmological constant, $\Lambda=-6/L^2_{AdS}$, where we set $L_{AdS}=1$.} will be linearized around the background solution of \cite{cy1}. Note that there is no background counterpart for the functions $C, D$ and $G$. The value of $k$ must be fixed from the beginning.

These linearized equations are too long to be reproduced here but they can be found in \cite{code}. In order to be able to organize the equations in a favorable structure, it is important to write them in a fully covariant way using derivatives along outgoing null rays $\dot h$, as defined in
\be
\dot h \equiv \partial_t h + \frac{1}{2} A\, \partial_r h\;, \quad d_3 h \equiv \partial_y h - F \partial_r h \;.
\label{derivdefs}
\ee
Note that $d_3 h$ is a derivative in the longitudinal direction orthogonal to radial null geodesics. Taking into account the previous decomposition into background and fluctuations, these definitions apply to the fluctuations as
\bea
\nonumber \dot{\delta h} &=& \partial_t (\delta h) + \frac{1}{2} A\, \partial_r (\delta h)+\frac{1}{2} \delta A\, \partial_r h_0\; ,
\\ d_3 (\delta h) &=& \partial_y (\delta h) - F\, \partial_r (\delta h) - \delta F\, \partial_r h_0 \;.
\label{derivdefs2}
\eea
The asymptotic analysis of Einstein's equations near the boundary provides a large $r$ expansion, where we include the extra gauge freedom $\xi$ described above, of the form
\begin{subequations}
\begin{flalign}
    A = (r+\xi)^2 -2 \partial_t \xi + \frac {a_4 +e^{i k x_1} \delta a_4 }{r^{2}} &+ O(r^{-3}) \, ,
\label{Aexp}
\\
    F = \partial_y \xi + \frac {f_4 +e^{i k x_1} \delta f_4}{r^2} &+ O(r^{-3})\,,
\label{Fexp}
\\
    G = e^{i k x_1}\frac { \delta g_4}{r^2} &+ O(r^{-3})\,,
\label{Gexp}
\\
    B = \frac{b_4+e^{i k x_1} \delta b_4}{r^{4}} &+ O(r^{-5})\,,
\label{Bexp}
\\
   C = e^{i k x_1} \frac{\delta c_4}{r^{4}} + O(r^{-5}), \; D = e^{i k x_1} \frac{\delta d_4}{r^{4}} &+ O(r^{-5})\,,
\label{CDexp}
\\
    \Sigma = r + \xi &+ O(r^{-7})\,,
\label{Sigmaexp}
\end{flalign}
\label{asymptotics}
\end{subequations}
We identify $a_4, \delta a_4, b_4, \delta b_4, \delta c_4, \delta d_4, f_4, \delta f_4, \delta g_4$ as the normalizable modes which are related to the stress-energy tensor of the dual theory. They are functions of ($t, y$), and they are not completely independent, since the previous expansions solve the equations only as long as the background coefficients satisfy
\be
\partial_t a_4 = -\tfrac{4}{3}\, \partial_y f_4 \,, \quad
\partial_t f_4 = -\tfrac{1}{4}\partial_y a_4 -2 \partial_y b_4\,,
\label{bdryevol}
\ee
and the inhomogeneities' coefficients satisfy
\bea
\nonumber \partial_t \delta a_4 &=& -\tfrac{4}{3}\, \left( \partial_y \delta f_4 + ik\, \delta g_4  \right) \,,
\\  \partial_t \delta f_4 &=& -\tfrac{1}{4}\partial_y \delta a_4 -2 \partial_y \delta b_4 + \partial_y \delta c_4+ ik\, \delta d_4 \,,
 \\ \partial_t \delta g_4 &=& -\tfrac{1}{4}ik\, \delta a_4 +ik\, \delta b_4 -ik\, \delta c_4+ \partial_y \delta d_4 \,. \nonumber
\label{bdryevol2}
\eea
These equations will be used to evolve the boundary conditions forward in time. They can equivalently be derived without making use of Einstein's equations, but from the conservation equations of the stress-energy tensor $\nabla^\mu T_{\mu\nu} =0$. They have a physical interpretation in terms of continuity conditions for the transport of energy and momentum.

Specifically, the expectation value of the stress-energy tensor of this problem contains energy density, momentum densities, pressures, and shear stress terms, all holographically mapped via gauge/gravity duality:
\be
 \langle T_{\mu\nu} \rangle = \frac{\nc^2}{2\pi^2} \begin{pmatrix}\mathcal{E} & \mathcal{S}_y& \mathcal{S}_{x_1} & 0 \\[1ex] \mathcal{S}_y & \mathcal{P}_y & \mathcal{T} & 0\\[1ex] \mathcal{S}_{x_1} & \mathcal{T} & \mathcal{P}_{x_1} & 0 \\[1ex] 0& 0 & 0 & \mathcal{P}_{x_2} \end{pmatrix} \,.
\ee
After transforming the asymptotic expansions (\ref{asymptotics}) to Fefferman-Graham coordinates, we can use the holographic renormalization prescription to extract the relations
\begin{subequations}
\begin{align}
\mathcal {E} &= - \tfrac{3}{4} (a_4 + \delta a_4) \,, \quad \mathcal {S}_y = f_4 + \delta f_4\,,  \\
& \mathcal {P}_y = - \tfrac{1}{4} (a_4 + \delta a_4) -2 (b_4+\delta b_4) + \delta c_4 \,, \\
\hspace{-1cm} \mathcal {S}_{x_1} &= \delta g_4 \,, \;\;\; \mathcal {P}_{x_1} =   - \tfrac{1}{4} (a_4+\delta a_4) + b_4 +\delta b_4- \delta c_4 \,, \\
\mathcal{T} &= \delta d_4 \,, \;\;\; \mathcal {P}_{x_2} =  - \tfrac{1}{4} (a_4+\delta a_4) + b_4 +\delta b_4 \,,
\end{align}
\label{setensorshocks}
\end{subequations}
where we have omitted the $e^{ik\, x_1}$ factors in front of every $\delta$ term.

\section{Numerics Overview}
\label{num}
A generic description of the numerical approach that can be applied to solve the dynamics of this problem is found in \cite{cyreview}.
By applying the characteristic formulation within AdS,
the set of coupled partial differential equations of GR can be very conveniently written as a nested set of linear ordinary differential equations.

\begin{figure}[b]
\begin{center}
\includegraphics[scale=0.7]{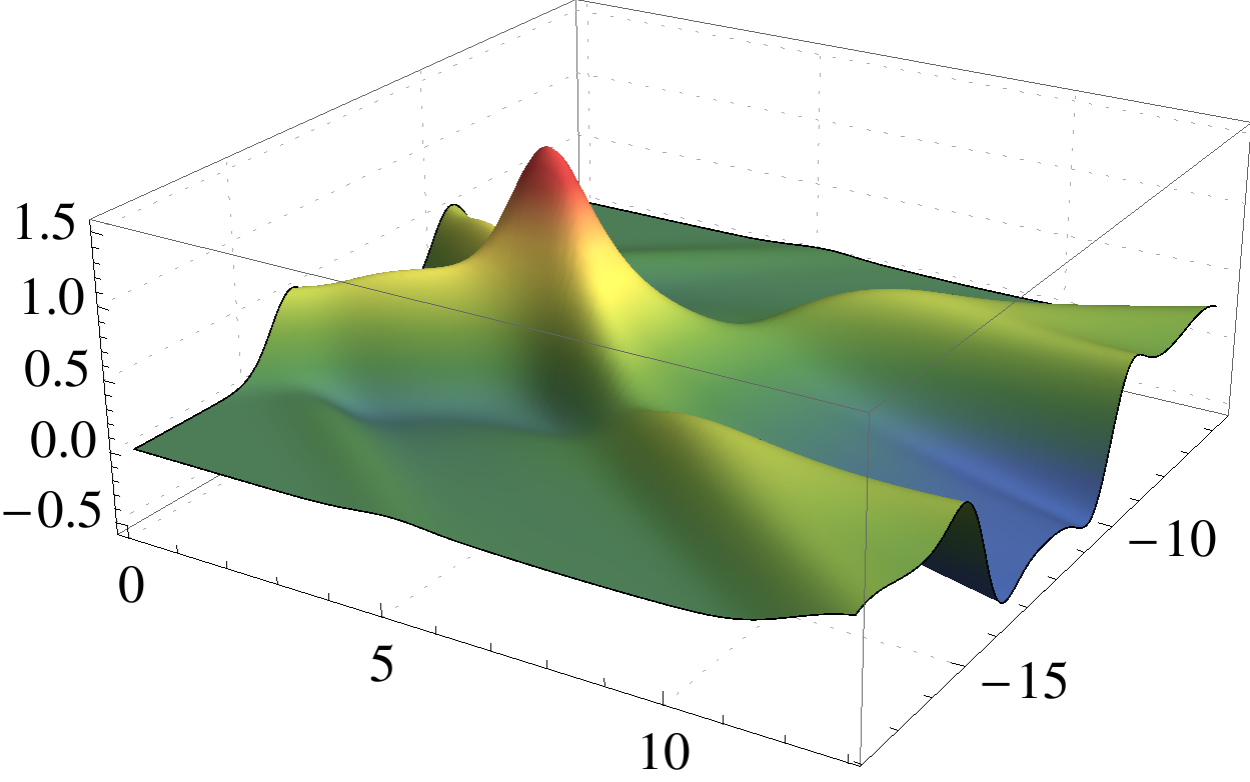}
\put(-160,6){\large{$\mu$t}}
\put(-27,30){\large{$\mu$y}}
\put(-230,130){\large{$\delta \mathcal E$}}
\caption{Inhomogeneity on the energy density, $\delta \mathcal E$ (from $\delta a_4$), for $k=0.2$,
as a function of longitudinal coord. $y$ and time $t$.}
\label{deltaa4}
 \end{center} 
\end{figure}

It is necessary to specify the spatial part of the metric on the initial time slice, except for the determinant (that is, except for $\Sigma$). Thus, one starts with the initial data provided for $\mathcal{F}_{in}=\{B_0, \delta B, \delta C, \delta D \}$. From there, the nested structure allows to solve for the other functions step by step, following the sequence
\be
\nonumber
\mathcal{F}_{in} \to S_0 \to F_0 \to \dot S_0 \to \dot B_0 \to A_0 \to \dot F_0 \to \delta S \to \delta F \to \delta G \to \dot{\delta S} \to \dot{\delta B} \to \dot{\delta C} \to \dot{\delta D} \to \delta A.
\ee

Note that the dotted functions are solved as if they were unrelated to their undotted counterparts. These 14 steps correspond to the 6 equations for the background, followed by the 8 linearized equations for the inhomogeneities. The numerical scheme that was implemented was based on pseudospectral methods, solving these equations at every time slice, then inverting~(\ref{derivdefs}-\ref{derivdefs2}) to obtain time derivatives and evolving the $\mathcal{F}_{in}$ forward to the next slice. An Adams-Bashforth method was carried out for this purpose. For many more details, see\cite{code}.

\begin{figure}[t]
\begin{center}
\includegraphics[scale=0.65]{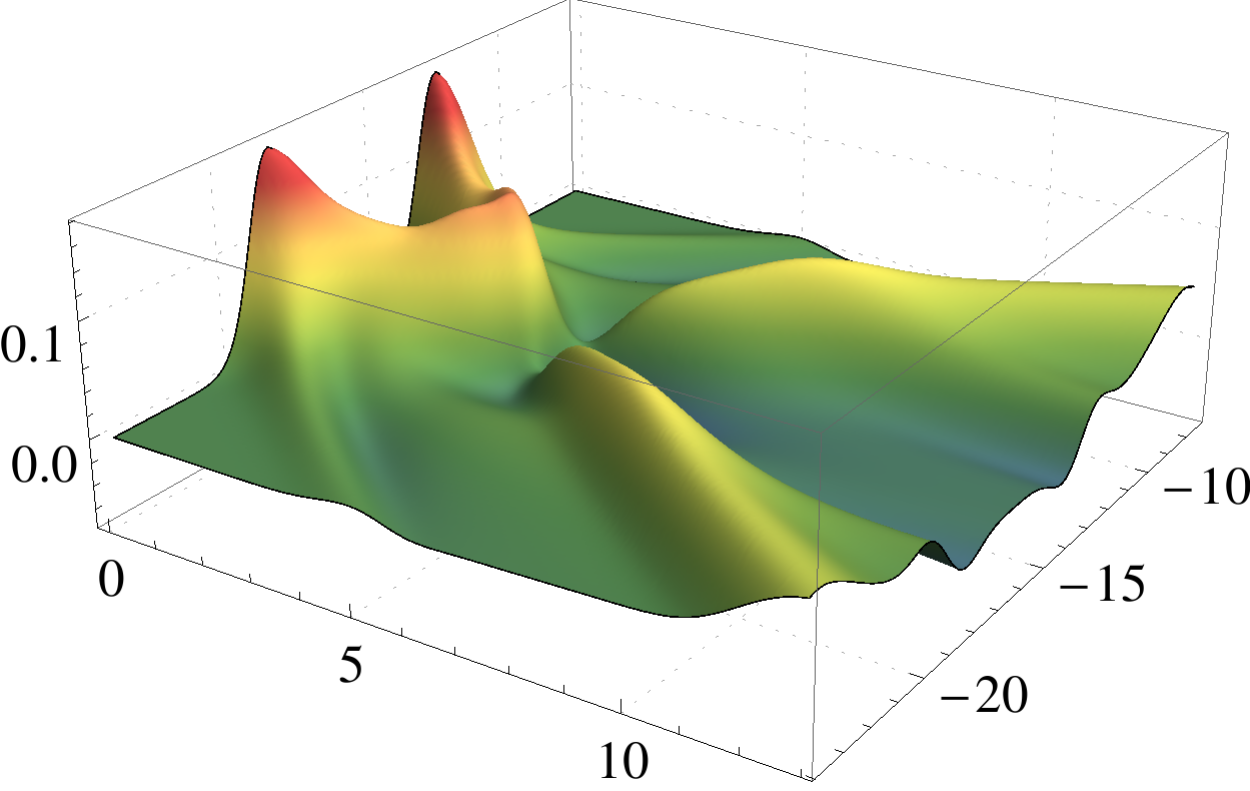}
\put(-160,6){\large{$\mu$t}}
\put(-27,20){\large{$\mu$y}}
\put(-220,120){\large{$\delta \mathcal P$}}
\caption{Inhomogeneity on the pressure anisotropy, defined by $\delta \mathcal P = \delta (\mathcal P_{x_1} + \mathcal P_y - 2 \mathcal P_{x_2})/3$, for $k=0.2$, as a function of longitudinal coordinate $y$ and time $t$.}
\label{deltab4}
 \end{center} 
\end{figure}

In addition, there are 4 constraints (equations that do not provide dynamics, but are evaluated to monitor accuracy): one is of the background and includes $\partial_t \dot S$, while the other three include $\partial_t (\dot{\delta S})$,  $\partial_r (\dot{\delta F})$ and  $\partial_r (\dot{\delta G})$ respectively.
Their asymptotic analysis near the boundary provides the conditions~(\ref{bdryevol}-\ref{bdryevol2}), which must be imposed to provide boundary conditions as the time evolution goes along. 
Given their self-fulfilling nature, the constraints can be left out of the calculation, only to be used as a convergence check of the numerics.

The initial data of the evolution is extracted from the metric of two planar shocks moving towards each other. For the calculations presented here, the same shocks as in \cite{cy1} were considered,
$\mathcal{H}(t,y) \equiv \frac{\mu^3}{\sqrt{2\pi w^2}} e^{-(t \mp y)^2 / 2w^2}\,, $
with $w=0.75/\mu$. We also chose a background energy density $\delta=0.075\mu^4$. This provides initial data for $B_0(t=0,r,y)$, and for the coefficients $a_4(t=0,y)$ and $f_4(t=0,y)$. But now this must be supplemented with initial data for the perturbations. In principle, any initial state can be considered, as long as Einstein's equations are satisfied (the inhomogeneities may take any shape).

In our calculations, we simply chose $\delta a_4, \delta f_4, \delta g_4$ so that the inhomogeneity behaves like a planar wave proportional to the amplitude of the background at each point, that is, $a_4 \to a_4(1+\epsilon e^{ik\,x_1})$, $f_4 \to f_4(1+\epsilon e^{ik\,x_1})$ and $\delta g_4 =0$. And for the bulk profiles of $\delta B, \delta C, \delta D$, we chose them to be given by the first terms in their respective expansions~(\ref{asymptotics}). This fixes the radial dependence and the boundary values are determined by~(\ref{bdryevol2}) so that
\be
\hspace{-0.2cm} \delta B (0,r,y) = \frac{a_4 (0,y)}{4 r^4} ,\; \delta C(0,r,y)=\delta D(0,r,y)=0\,.
\label{ic}
\ee
For these initial conditions, functions $\delta C$, $\delta D$ and $\delta G$ acquire non-vanishing profiles spontaneously.
It would be interesting to study different choices of initial conditions, in order to check how they affect the stability of the propagation of the shocks before the collision, but we choose to leave such an analysis to future work.

\begin{figure}[b]
\begin{center}
\includegraphics[scale=0.7]{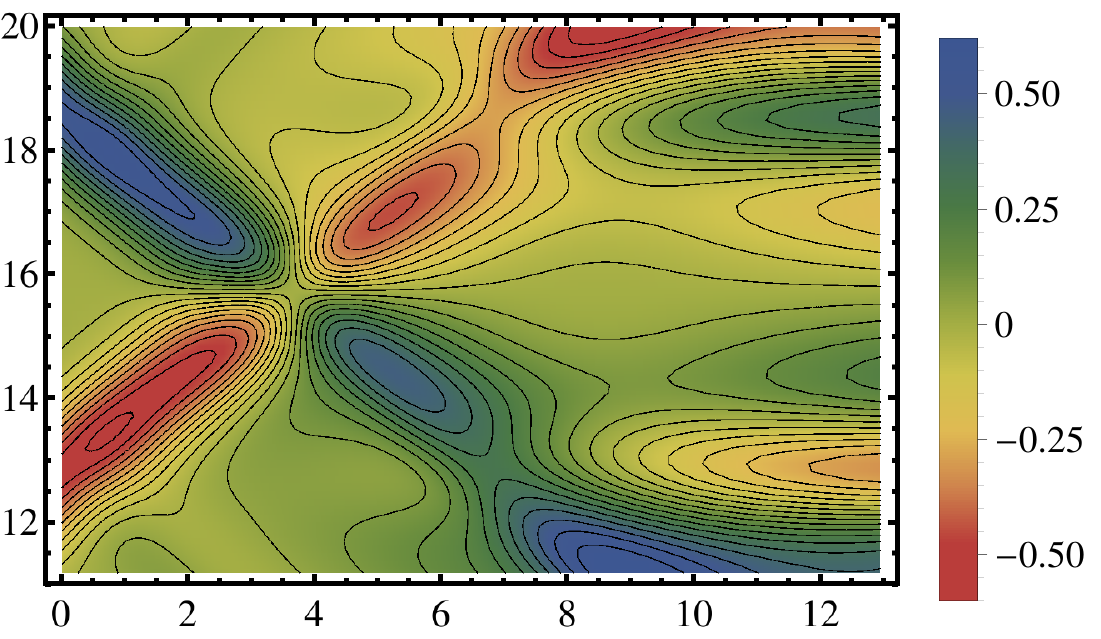}
\put(-47,-3){\normalsize{$\mu$t}}
\put(-232,110){\normalsize{$\mu$y}}
\caption{Inhomogeneity on the longitudinal energy flux, $\delta \mathcal S_y$, for $k=0.5$, as a function of longitudinal coord. $y$ and time $t$.}
\label{deltaf4}
\end{center}
\end{figure}

Several inhomogeneities of the expected stress-energy tensor are plotted in Figs.~(\ref{deltaa4}-\ref{deltaf4}). Note that since the equations are linearized, the overall amplitude of these inhomogeneities is completely irrelevant. Their sign is also irrelevant, since each of these figures corresponds to an $x_1=\text{ctant.}$ slice, and they oscillate along the transverse direcion. This is due to the factor $e^{i k x_1}$ they bear in front.

\section{Apparent Horizon}
\label{ah}
As discussed in \cite{cyreview},the residual gauge freedom $r \to r+\xi(t,y)$,
is fixed by imposing the AH to lie at a constant $r$, for instance $r=1$. This is easily carried out by absorbing any deviation into the chosen gauge, that is, $\delta \xi = \rah-1$. The computation behind is a crucial part of the numerical calculation, so it merits to give a further explanation about this.

To find the position of the AH, $\rah(y)$, \cite{cy1} gives
\be
\hspace{-0.06cm} \left. 3 \Sigma^2 \, \dot \Sigma - \partial_y (e^{2B}\, F \, \Sigma) + {\tfrac{3}{2}} e^{2B} \, F^2\,\partial_r \Sigma \right|_{r=\rah(y)} = 0\;,
\label{ah}
\ee
where the functions here can be those of the background (the inhomogeneities make a negligible contribution). This equation can be solved by finding the root of that expression. However, it is derived under the assumption that the AH lie at a constant position $r=\rah$, instead of a trajectory $r=\rah(y)$. Otherwise,~(\ref{ah}) must be modified by inverting the gauge transformation that would have led us to having it at a constant position in the first place. As can be seen in its expansion~(\ref{Fexp}), this entails the explicit change $F \to F - \partial_y \xi$, as well as evaluating every function at $r+\xi$. This is a significant complication, since the problem becomes an intricate non-linear differential equation for $\rah(y)$ (or, equivalently, $\xi(y)$).

Given a time slice $t_0$,~(\ref{ah}) gives the correct AH after performing an iterative procedure to find $\xi(t_0,y)$. But during the time evolution, it is more efficient to demand the time derivative to vanish,
\be
\partial_t \left( 3 \Sigma^2 \, \dot \Sigma - \partial_y (e^{2B}\, F \, \Sigma) + {\tfrac{3}{2}} e^{2B} \, F^2\,\partial_r \Sigma \right) = 0\;.
\ee
It can be seen~\cite{code} that this constitutes a linear $2^{\text{nd}}$ order differential equation for $\partial_t \xi$, so its time evolution can be readily computed. But, since we are using the equation that assumes a constant $\rah$, this approach introduces some error in the calculation, which must be corrected by performing the explicit calculation of $\xi(t_0,y)$ occasionally during the evolution (every 10-20 timesteps).

\section{Hydrodynamics}
\label{hydro}

Comparing our results with those for the background (see \cite{cy1}), a common observation can be drawn: the presence of the inhomogeneities persists even after the background has equilibrated. For instance, we can see that the energy density $\mathcal E$ is spread out by the time $t\sim 10/\mu$. However, we find that there is still a very uneven profile in $\delta \mathcal E$ well after that time, as shown in fig.~(\ref{deltaa4}).
This is an interesting observation, because the presence of such a perturbation in the energy density at later times would give some non-trivial structure to the otherwise flat spatial profile of the local energy density \cite{cwds}.

However, it is important to keep in mind that the program of research on holographic thermalization has created two well established concepts of what should be understood as thermalization. On the one hand, one could refer to the isotropization of the stress tensor in the local rest frame. This can take a very long time\footnote{Whenever a time is refered to as ``long'' or ``short'' in the context of the QGP, it is understood to be with respect to the scale set by the local temperature.}. On the other hand, one could refer to the applicability of viscous hydrodynamic constitutive relations. This is sometimes called ``hydrodynamization'', and the consensus in this respect is that it is surprisingly short. On the following, we shall refer to the latter definition of thermalization.

\begin{figure}[t]
\begin{minipage}{0.47\textwidth}
\begin{center}
\includegraphics[scale=0.65]{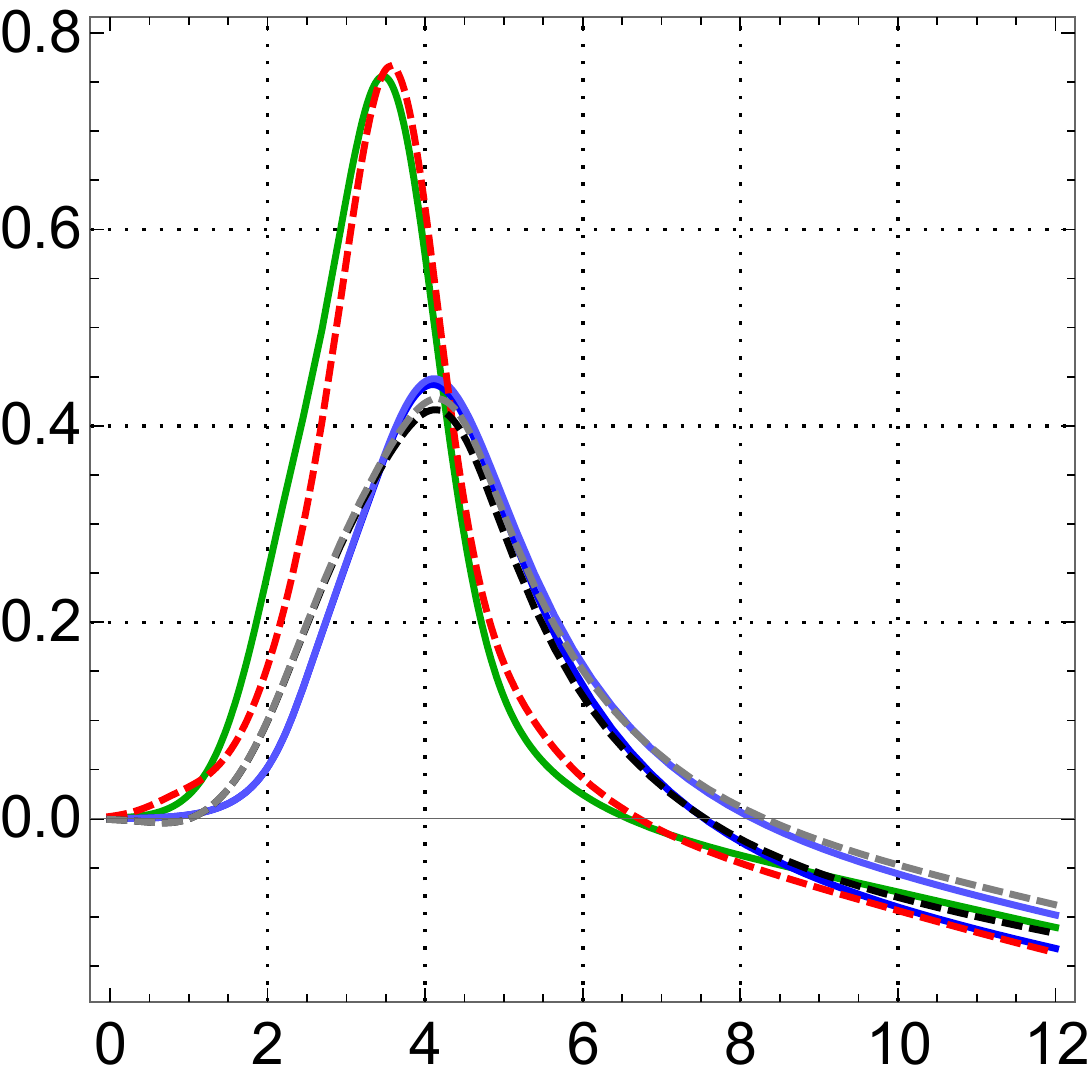}
\put(-23,-9){\normalsize{$\mu$t}}
\put(-225,175){\normalsize{$\delta \mathcal P/\mu^4$}}
\end{center}
\end{minipage}
\hfill
\begin{minipage}{0.47\textwidth}
\begin{center}
\includegraphics[scale=0.65]{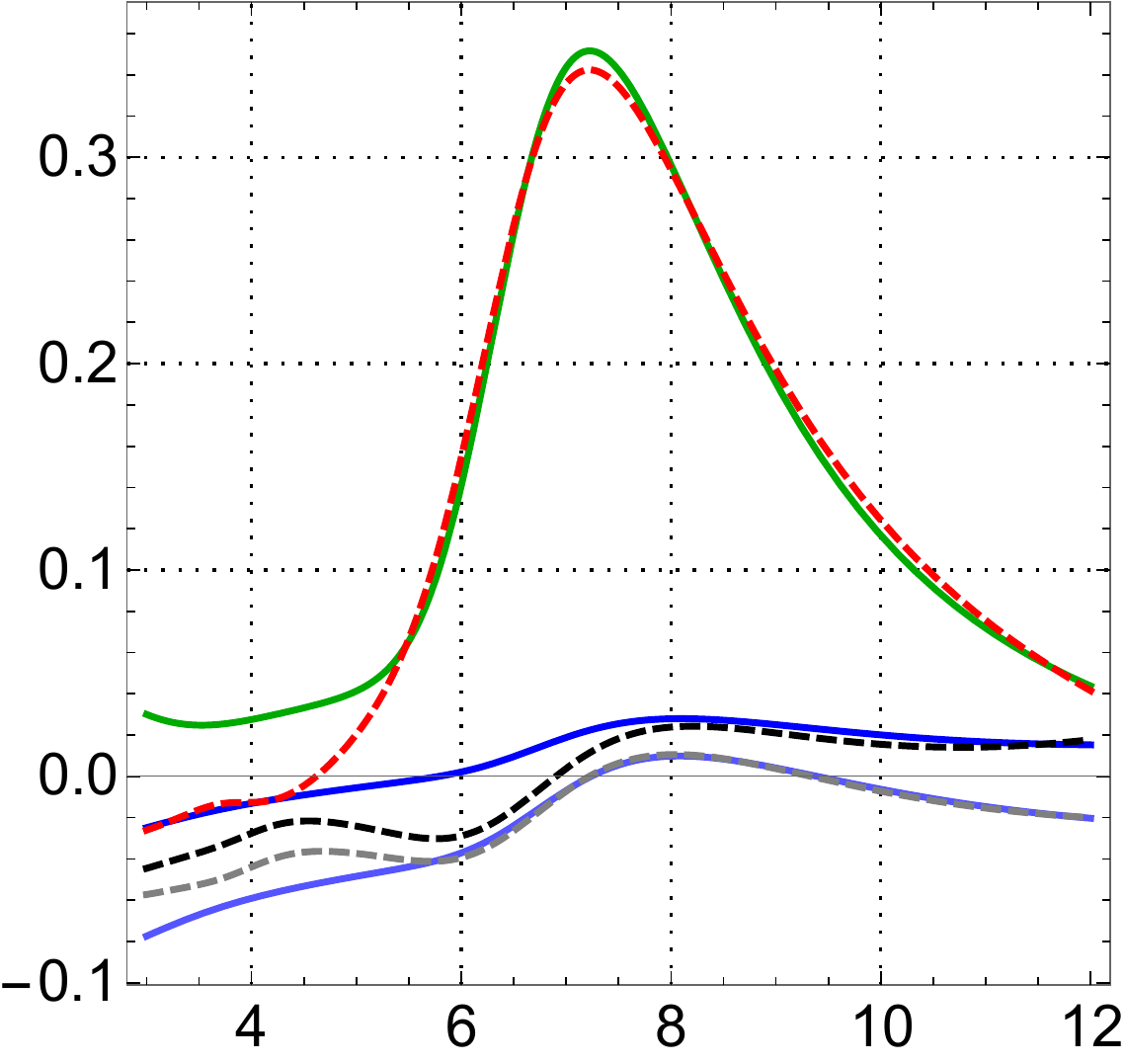}
\put(-23,-9){\normalsize{$\mu$t}}
\put(-225,190){\normalsize{$\delta \mathcal P/\mu^4$}}
\end{center}
\end{minipage}
\label{hydroplot}
\caption{Inhomogeneities in the longitudinal (green) and transverse (blue) pressures, for $k=0.5$, as a function of time, at mid-rapidity ($y=y_0$) and at $y=y_0+3/\mu$. Also shown for comparison are the corresponding contributions to the pressures predicted by the viscous hydrodynamic constitutive relations (dashed lines). At late times, each dashed line overlaps with its corresponding solid line.}
\end{figure}

Thus, in order to draw comparisons with previous work, we need to test the validity of hydrodynamics. To do so, we compare the actual pressures from the boundary stress-energy tensor with the pressures that would follow if the viscous hydrodynamic constitutive relations were satisfied \cite{hydropaper1,hydropaper2}. Results from such a comparison can be found In Fig.~(\ref{hydroplot}), where we plot the inhomogeneities in the longitudinal $\delta \mathcal P_y$ and transverse $\{\delta \mathcal P_{x_1}, \delta \mathcal P_{x_2}\}$ pressures as a function of time, at two specific points, $y=y_0=5\pi/\mu$ (this is the point where our shocks actually collide) and $y=y_0+3/\mu$. The dashed lines show the fluctuations in the pressures $\delta \mathcal P^{\text{{\tiny hydro}}}$ as predicted by the hydrodynamic equations. 

The agreement with hydrodynamics is quite remarkable. In particular, at $y=y_0$, the sudden increase in the pressures due to the collision is reflected in a dramatic rise in their inhomogeneities, as expected. During this stage, the system is very anisotropic and far from equilibrium, and as a consequence hydrodynamics is not expected to be applicable (as reflected in \cite{cy1} for the background). Surprisingly, the hydrodynamic constitutive relations seem to hold almost from the beginning for the fluctuations considered here.

At late times, the pressure inhomogeneities asymptotically approach each other. This process of isotropization has not been completed even at $t=12/\mu$, which is much larger than the time it takes for the background to become isotropic. Thus, our fluctuations provide a nice example of a system where isotropization time and the hydrodynamization time are completely different.

\section{Discussion}
\label{discussion}
The transverse dynamics in a collision of gravitational shockwaves in AdS has been previously considered in \cite{wradial}, where the longitudinal dynamics were approximated as boost-invariant, and rotational symmetry is assumed. The dynamics we have discussed here do not require boost-invariance and describe the behavior of inhomogeneities in the shockwave which are allowed to propagate in a transverse direction. Our main approximations are the linearization of the equations of motion~(\ref{planewaves}), and the simple choice of initial conditions~(\ref{ic}).

The results apply qualitatively to all strongly coupled 4D conformal gauge theories with a gravitational dual description.
Typically, a closer approach to QCD requires introducing quarks, or fundamental matter. However, the gluons are the dominant degrees of freedom at the timescales involved in a heavy ion collision, and this is what motivates the idealization of considering nothing more than pure gravity. Still, tt is not obvious a priori how good an approximation to the QGP dynamics it can provide. As usual, eventual comparison with experimental observations is what can support the assumptions taken in holographic models.

At the experiments of the RHIC and LHC, density perturbations may exist and play a relevant role in the results. In order for the holographic approaches to make contact with this, inhomogeneities need to be included. Our disposal~(\ref{planewaves}) is limited, but also a first step towards a complete calculation. A different but also relevant phenomenon is the presence of radial flow, for which an expanding fireball of finite size would need to be considered, as opposed to a wave of infinite transverse extent. In order to model this, one would need to go beyond the fluctuation approximation.

Based on the typical behavior manifested by our fluctuations (as shown in Fig.~(\ref{hydroplot})), we infer that one is to expect inhomogeneous profiles to keep being inhomogeneous for a relatively long time. Quite longer than the thermalization time, which would be unaffected by the inhomogeneities, given that they experience a very fast hydrodynamization. This result supports the use of hydrodynamic approximations to study the propagation of perturbations during the out-of-equilibrium stage of QGP formation. However, one should note that the agreement observed in our results can be related to the linearized treatment we have used to solve the equations.

Furthermore, interesting information could be extracted from a thorough analysis of the results. In particular, considering different initial configurations for the inhomogeneities would allow to establish which perturbations are physically relevant and which are not.
It would also be possible to see at which point of the evolution do they become hydrodynamical, or whether they do at all.
Additionally, a byproduct of this calculation would be the spectrum of quasinormal modes for finite $k$ of the final black hole. A hydrodynamic gradient expansion of the dynamics could allow to read them off \cite{qnmheller}. We leave this to future work.

Another estimulating extension of this project would be to allow for an interaction between modes with different momenta, since in that case turbulent effects may arise. In recent years, it has been discovered that turbulence is an ubiquitous property of Gravity \cite{cheslerturbulent, lehner}. However, if symmetries are forced into the system, turbulence is missed.
Intuitively, one can argue that turbulent effects would lead to a shorter thermalization time, due to the cascading behavior towards higher modes (both in the transverse and longitudinal directions).

Finally, it is possible that turbulence may arise in the calculation presented here at later times, since the perturbations propagate in a non-trivial time dependent background. The perturbed mode could resonate with its pattern. This could be an appealing analysis too.

\begin{acknowledgments}
I would like to thank Elias Kiritsis, Aristomenis Donos, David Mateos, Luis Lehner, Amos Yarom, Federico Carrasco and Miquel Triana for valuable discussions, and especially Wilke van der Schee for his help with the technical aspects of the calculation.
I am also grateful to DAMTP at the University of Cambridge, the Technion of Israel and Paris ENS for hospitality.
The author is supported in part by E.U.'s $7^{\text{th}}$ Framework Programme under grant agreements (FP7- REGPOT-2012-2013-1) number 316165, PIF-GA-2011- 300984, the E.U. program \emph{``Thales'}' and \emph{``Herakleitos II"} ESF/NSRF 2007-2013 and co-financed by the E.U. (European Social Fund, ESF) and Greek national funds through the Operational Program \emph{``Education and Lifelong Learning''} of the National Strategic Reference Framework (NSRF) under \emph{``Funding of proposals that have received a positive evaluation in the $3^{rd}$ and $4^{th}$ Call of ERC Grant Schemes''}.
\end{acknowledgments}

\end{document}